\documentstyle[12pt]{article}

\def\be{\begin{equation}} 
\def\ee{\end{equation}}
\def\bea{\begin{eqnarray}} 
\def\eea{\end{eqnarray}}

\begin{document} 
\title{ \vspace{-2cm}\flushright{\small IP/BBSR/96-45 \\ hep-th/9609015} \\
\vspace{2cm}On Orientifold Constructions of Type IIA Dual Pairs}
\author{Anindya K. Biswas\thanks{akb@iopb.ernet.in} \\
 Institute of Physics, \\Bhubaneswar 751 005, INDIA} 
\date{\today}
\maketitle
\begin{abstract}
\noindent 
In this paper we analyze the earlier constructions of the 
type IIA dual pairs through orientifolding. 
By an appropriate choice of $\Gamma$-matrix basis for 
the spinor representations of the $U$-duality group,
we give an explicit relationship between 
the orientifold models and their dual pairs. 
\end{abstract}

The constructions of ``strong-weak'' dual pair \cite{witt2} 
of string
theories, and in particular the construction of type II dual pair
of string theories in various dimensions have been investigated 
by several authors,\cite{witt1,seva,Kumar}.
 In these constructions, the $SO(5, 5; Z)$
U-duality symmetry of the six dimensional type II string theory 
was used to construct several dual pair of string theories upon 
further compactifications of the extra dimensions. The matching
of the massless spectra in four dimensions was observed as an 
evidence that these theories are indeed dual to each other. 
Other type II dual pair models include the orientifold examples
in four dimensions\cite{Kumar} and two dimensional examples\cite{sen}
. Recently a
$U$-duality invariant partition function for these models have 
been proposed which gives the degeneracy of the fundamental as well
as the solitonic states\cite{dvv}. \\

In an earlier paper, a class of type II dual pair models were 
constructed through orientifolding\cite{Kumar}.
 It was found that the type IIA
string theory in ten dimensions 
possesses certain discrete symmetries which 
cannot be embedded into the $T$-duality group in lower 
dimensions. However, it was argued that two such discrete symmetries
 are related by ``strong-weak'' duality and 
can be used for model building. In this paper we examine
the relationship among these discrete symmetries in detail. 
This is achieved by arranging the fields into appropriate 
representations of the $U$-duality group viz. $SO(5, 5; Z)$ 
and finding their transformations. 
Interestingly, we find that in the present case the 
$U$-duality relates the 
discrete symmetries which are themsleves not the elements of the
$SO(5, 5)$ symmetry.
\par
Let us now start by briefly reviewing the discrete symmetries of 
the type IIA theory as discussed in \cite{Kumar}. The bosonic field 
content of the type IIA theory in ten dimensions are 
the graviton $G_{M N}$, antisymmetric tensor $B_{M N}$, dilaton
$\phi$, the one-form field $A_{M}$ and the three-form field $C_{M N P}$.
Among these, the gaviton, dilaton and the antisymetric tensor fields
arise from the NS-NS sector of the theory and the one and three-form 
fields from the R-R sector.
This theory has several discrete symmetries. The ones
known as ``T-duality'' have been studied extensively. One can,
however, show that there are other discrete symmetries present as 
well. One such symmetry referred to as $Z_{2}^{0}$ is used for
constructing orientifold models. Under this, string worldsheet
changes its orientation. At the same time, there is a change
 of sign for the coordinates, $(X^{5},...,X^{9})$. Another 
symmetry, relevant for us in this paper is an improper $T$-
duality rotation. This will be referred to as $Z_{2}^{\star}$
and acts as $(X^{5},..,X^{8})\rightarrow-(X^{5},..,X^{8})$.
Both $Z_{2}^{0}$ and $Z_{2}^{\star}$ remain the symmetries of
the type IIA action upon toroidal compactifications. In particular,
for the compactifications to four dimensions and on the basis of 
strong analogy with the results of \cite{seva}, it was pointed out
in \cite{Kumar} that by modding out the original theory by these
discrete symmetries leads to models which are ``strong-weak''
dual pair.

We now follow the general approach of references \cite{seva,Kumar} and 
show that the two discrete symmetries discussed in the last 
paragraph are connected by a $U$-duality symmetry.
The particular $U$-duality element which connects these
is, however, different than that in \cite{Kumar} by a $T$-duality 
factor. This is still, however, an element of $SO(5,5)$ and
has a ten dimensional matrix representation.
We now obtain 
this representation by arranging the ten dimensional fields
in terms of $U$-duality multiplets  $
Q_{{\bar\mu}\bar{\nu}\bar{\rho}}$
, $\tilde{M}$ and $P_{{\bar\mu}}$ as: 

\bea 
Q_{{\bar\mu}{\bar\nu}{\bar\rho}}
\equiv\pmatrix{H_{{\bar\mu}{\bar\nu}{\bar\rho}} \cr
                         e^{-2\phi}\tilde{H}_{{\bar\mu}\bar{\nu}\bar{\rho}}\cr 
                          F_{{\bar\mu}\bar{\nu}\bar{\rho}6}\cr
                          F_{{\bar\mu}\bar{\nu}\bar{\rho}7}\cr
                          F_{{\bar\mu}\bar{\nu}\bar{\rho}8}\cr
                          F_{{\bar\mu}\bar{\nu}\bar{\rho}9}\cr
                      F^{'}_{{\bar\mu}\bar{\nu}\bar{\rho}6}\cr
                      F^{'}_{{\bar\mu}\bar{\nu}\bar{\rho}7}\cr
                      F^{'}_{{\bar\mu}\bar{\nu}\bar{\rho}8}\cr
                      F^{'}_{{\bar\mu}\bar{\nu}\bar{\rho}9}}
\equiv\pmatrix{H_{{\bar\mu}{\bar\nu}{\bar\rho}} \cr
		e^{-2\phi} \tilde{H}_{{\bar\mu}{\bar\nu}{\bar\rho}}\cr
		\vec{D}_{{\bar\mu}{\bar\nu}\bar{\rho}}},\nonumber\\
\noindent	
\tilde{M}=\pmatrix{e^{2\phi}& -1/2e^{2\phi}\psi^{T}L\psi&-e^{2\phi}\psi^{T}\cr
                                                                   \cr
      &e^{-2\phi}+\psi^{T}LR_{s}(M)L\psi & \psi^{T}LR_{s}(M)\cr
  & +1/4e^{2\phi}(\psi^{T}L\psi)^{2} & +1/2e^{2\phi}\psi^{T}(\psi^{T}L\psi)\cr
                                                                      \cr
  &               &R_{s}(M)+e^{2\phi}\psi\psi^{T} },\nonumber
P_{{\bar\mu}}=\pmatrix{ C_{{\bar\mu}69}\cr
                         C_{{\bar\mu}79}\cr
                         C_{{\bar\mu}89}\cr
                         K^{'}_{{\bar\mu}}\cr
                         C_{{\bar\mu}78}\cr   
                         C_{{\bar\mu}68}\cr 
                         C_{{\bar\mu}67}\cr
                         A_{{\bar\mu}}\cr
                         g_{{\bar\mu}6}\cr
                         g_{{\bar\mu}7}\cr
                         g_{{\bar\mu}8}\cr
                         g_{{\bar\mu}9}\cr                     
                         B_{{\bar\mu}6}\cr
                         B_{{\bar\mu}7}\cr
                         B_{{\bar\mu}8}\cr
                         B_{{\bar\mu}9}},\nonumber\\
\eea

\noindent
where F and $F^{'}$ in the expression
for Q are the original and dual field strengths
of the six dimensional RR 3-form field C. These field strengths 
are related by 
 $\tilde{Q}
=\tilde{M}\tilde{L}\,Q +O(\psi)$ 
\cite{seva}. Hence to zeroeth order in $\psi$, $H$ and 
$\tilde{H}$ are Poincare dual.\quad
In eq.(1) only upper triangular entries of 
$\tilde{M}$ has been written, as this is a symmetric matrix. 
$\tilde{M}$ contains $M$ and $\psi$. $M$ is  
defined in terms of toroidal
moduli g, B and $\psi$ in terms of various RR scalars. 
They are given as,
\noindent
\bea
M=\pmatrix{ g^{-1} & -g^{-1}B\cr
           Bg^{-1} & g-Bg^{-1}B },\nonumber\quad
g=\pmatrix{g_{mn} & g_{m9}\cr
           g_{9n} & g_{99}},\nonumber\quad
B=\pmatrix{B_{mn} & B_{m9}\cr
           B_{9n} & 0},\nonumber\quad
\psi=\pmatrix{A_{6}\cr
              .    \cr
              .    \cr
              A_{9}\cr
             C_{789}\cr 
             C_{689}\cr 
             C_{769}\cr
             C_{678}},
\eea
where; m,n = 6,., 8.
\noindent
$\tilde{M}$ is an $SO(5,5)$ matrix i.e. it
satisfies the relation,\\
$\tilde{M}\tilde{L}\tilde{M}=\tilde{L}$ where,
\noindent
\be
\tilde{L}=\pmatrix{ \sigma_{1} &0 \cr
                       0  &\tilde{L}_{8} },\quad
\tilde{L}_{8}=\pmatrix{ & 0&I_{4}\cr
                          &I_{4}& 0 }.
\ee
\noindent
In addition, we have 
$g_{55}$ and the graviton 
$g_{\mu \nu}$. 

Then symmetries defined as  
$Z_{2}^{0}$ and $Z_{2}^{\star}$ do not transform   
the fields $g_{\mu \nu}$, $g_{55}$
as well as dilaton $\phi$.
For describing other transformations,
in the following, we will use 
barred(unbarred) indices for six(five) dimensional
coordinates. As our objective will be to relate
the two discrete symmetries we consider $Z_{2}^{0}$
 and $Z_{2}^{\star}$ operations on 
 two separate multiplets rather
 than on the same multiplet. We will denote the two 
 multiplets as primed 
 and unprimed respectively. We will add superscripts
 $0$ and $\star$ respectively 
 on them to denote transformed multiplets.
\noindent
 $Z_{2}^{0}$ transforms unprimed multiplets
  as follows:\\
\bea
Q_{{\bar\mu}{\bar\nu}\bar{\rho}}^{0}
          =\pmatrix{-H_{\mu\nu\rho},&H_{\mu\nu5} \cr
           e^{-2\phi}\tilde{H}_{\mu\nu\rho},&-e^{-2\phi}\tilde{H}_{\mu\nu5}\cr
                   F_{\mu\nu\rho6},&-F_{\mu\nu56} \cr                                             F_{\mu\nu\rho7},&-F_{\mu\nu57} \cr
                    F_{\mu\nu\rho8},&-F_{\mu\nu58} \cr
                    F_{\mu\nu\rho9},&-F_{\mu\nu59} \cr
                    -F^{'}_{\mu\nu\rho6},&F^{'}_{\mu\nu56} \cr
                    -F^{'}_{\mu\nu\rho7},&F^{'}_{\mu\nu57} \cr
                    -F^{'}_{\mu\nu\rho8},&F^{'}_{\mu\nu58} \cr
                    -F^{'}_{\mu\nu\rho9},&F^{'}_{\mu\nu59}},\nonumber
\psi^{0}=\pmatrix{-A_{6}\cr
              -.    \cr
              -.    \cr
              -A_{9}\cr
             C_{789}\cr 
             C_{689}\cr 
             C_{769}\cr
             C_{678}},\nonumber 
P_{{\bar\mu}}^{0}=\pmatrix{- C_{\mu69},&C_{569}\cr
                        -C_{\mu79},&C_{579}\cr
                        -C_{\mu89},&C_{589}\cr
                        K^{'}_{\mu},&-K^{'}_{5}\cr
                        -C_{\mu78},&C_{578}\cr   
                        -C_{\mu68},&C_{568}\cr
                        -C_{\mu67},&C_{567}\cr
                        A_{\mu},&-A_{5}\cr   
                        -g_{\mu6},&g_{56}\cr        
                        -g_{\mu7},&g_{57}\cr 
                        -g_{\mu8},&g_{58}\cr
                        -g_{\mu9},&g_{59}\cr
                        B_{\mu6},&-B_{56}\cr
                        B_{\mu7},&-B_{57}\cr
                        B_{\mu8},&-B_{58}\cr
                        B_{\mu9},&-B_{59}}.\nonumber
\eea
\noindent
\be
g^{0}=g,\quad B^{0}=-B.
\ee
\noindent
We can rewrite the above transformations
in matrix form
as in Ref.\cite{seva}:
\bea
\noindent
Q_{{\bar\mu}{\bar\nu}\bar{\rho}}^{0}
\equiv \Omega(Z_{2}^{0})Q_{{\bar\mu}{\bar\nu}\bar{\rho}},\nonumber\quad
\tilde{M}^{0}\equiv \Omega(Z_{2}^{0})\tilde{M}\Omega(Z_{2}^{0})^{T},
\nonumber\quad
P_{{\bar\mu}}^{0}\equiv \tilde{R_{s}}(Z_{2}^{0})P_{{\bar\mu}},\nonumber
\eea
where,
\be
 \Omega(Z_{2}^{0})=\pmatrix{-1 & & & \cr
                               &1& & \cr
                               & &I_{4}& \cr
                               & &  &-I_{4}},\quad
  \tilde{R_{s}}(Z_{2}^{0})=\pmatrix{-I_{3}& & & & & \cr
                           &1& & & & \cr
                           & &-I_{3}& & & \cr
                           & & &1 & & \cr
                           & & &  &-I_{4}& \cr
                           & & &  &  &I_{4}}.\nonumber\\
\ee
\noindent
provided 
\be
R_{s}(\Omega^{0})=\Omega^{0}
\ee\\
where, $\Omega^{0} = \pmatrix{ I_{4} & \cr
                                 & -I_{4}}$.

\noindent 
Similarly the action of $Z_{2}^{\star}$ on primed multiplets are
as follows:
\bea
(Q_{{\bar\mu}{\bar\nu}\bar{\rho}}^{'})^{\star}
    =\pmatrix{H_{\mu\nu\rho},&-H_{\mu\nu5} \cr
           -e^{-2\phi}\tilde{H}_{\mu\nu\rho},&e^{-2\phi}\tilde{H}_{\mu\nu5}\cr
                   -F_{\mu\nu\rho6},&F_{\mu\nu56} \cr                                             -F_{\mu\nu\rho7},&F_{\mu\nu57} \cr
                    -F_{\mu\nu\rho8},&F_{\mu\nu58} \cr
                    F_{\mu\nu\rho9},&-F_{\mu\nu59} \cr
                    F^{'}_{\mu\nu\rho6},&-F^{'}_{\mu\nu56} \cr
                    F^{'}_{\mu\nu\rho7},&-F^{'}_{\mu\nu57} \cr
                    F^{'}_{\mu\nu\rho8},&-F^{'}_{\mu\nu58} \cr
                    -F^{'}_{\mu\nu\rho9},&F^{'}_{\mu\nu59}}^{'},\nonumber
(\psi^{'})^{\star}=\pmatrix{-A_{6}\cr
                           -.    \cr
                           -.    \cr
                            A_{9}\cr
                          C_{789}\cr 
                          C_{689}\cr 
                          C_{769}\cr
                         -C_{678}}^{'},\nonumber 
(P_{{\bar\mu}}^{'})^{\star}=\pmatrix{ -C_{\mu69},&C_{569}\cr
                        -C_{\mu79},&C_{579}\cr
                        -C_{\mu89},&C_{589}\cr
                        -K^{'}_{\mu},&K^{'}_{5}\cr
                        C_{\mu78},&-C_{578}\cr   
                        C_{\mu68},&-C_{568}\cr
                        C_{\mu67},&-C_{567}\cr
                        A_{\mu},&-A_{5}\cr   
                        -g_{\mu6},&g_{56}\cr        
                        -g_{\mu7},&g_{57}\cr 
                        -g_{\mu8},&g_{58}\cr
                         g_{\mu9},&-g_{59}\cr
                        -B_{\mu6},&B_{56}\cr
                        -B_{\mu7},&B_{57}\cr
                        -B_{\mu8},&B_{58}\cr 
                         B_{\mu9},&-B_{59}}^{'}.\nonumber
\eea            
\noindent
\be
g^{\star}=\pmatrix{g_{mn}& -g_{m9} \cr
                   -g_{9n}& g_{99}},\quad 
B^{\star}=\pmatrix{B_{mn} & -B_{m9}\cr
                  -B_{9n} & 0}.
\ee
\noindent 
Again we can rewrite the 
above transformations in the matrix form

$(Q_{{\bar\mu}{\bar\nu}\bar{\rho}}^{'})^{\star}\equiv
\Omega(Z_{2}^{\star})\,Q_{{\bar\mu}{\bar\nu}\bar{\rho}}^{'}$,\quad
$(\tilde{M}^{'})^{\star}\equiv
\Omega(Z_{2}^{\star})\,\tilde{M}^{'}\,\Omega(Z_{2}^{\star})^{T}$,\quad
$(P_{{\bar\mu}}^{'})^{\star}
\equiv\tilde{R_{s}}(Z_{2}^{\star})\,P_{{\bar\mu}}^{'}$ 

where
\be
 \Omega(Z_{2}^{\star})=\pmatrix{ 1 & & & & & \cr
                                         &-1&0&0&0&0 \cr
                                         &0&-I_{3}&0&0&0  \cr
                                         &0&0&1&0&0  \cr
                                         &0&0&0&I_{3}&0 \cr
                                         &0&0&0&0&-1},\nonumber
\tilde{R_{s}}(Z_{2}^{\star})=\pmatrix{-I_{4}& & & & & \cr
                                    &I_{4}& & & & \cr
                                    & & -I_{3}& & & \cr
                                    & &  &1 & &  \cr
                                    & &  &  &-I_{3}& \cr
                                    & &  &  &     &1 }
\ee
\noindent
provided once again
\be
R_{s}(\Omega^{\star}) = \Omega^{\star}
\ee\\
with $\Omega^{\star} = \pmatrix{-I_{3} & & & \cr
                                        &1& & \cr
                                        & &I_{3}& \cr
                                        & & &-1 }$.

Now let us verify eqns (3) and (5). We note
that $\Omega^{0}$,  $\Omega^{\star}$  and 
corresponding $R_{s}(\Omega)$'s, act on the
multiplet basises of Ref.\cite{Kumar}. They
are $SO(8)$ matrices rather than $SO(4,4)$ ones. We 
apply  $(\eta^{'})^{T}$  on them to transform 
to matrices acting on the multiplet basises of Ref.\cite{seva},
where $\eta^{'} = (1/\sqrt{2})\pmatrix{I_{4} & I_{4} \cr
                       -I_{4} & I_{4}}$.
Formally, we write the transformed matrices acting on 
the multiplet basises of Ref.\cite{seva} as $\Omega$ and $R_{s}$.
Then we embed them in $SO(10)$ representation as:
\be
\bar{R}_{s}(\bar{\Omega}) = \pmatrix{R_{c}(\Omega) & \cr
                                     & \Omega},\quad
\bar{\Omega} = \pmatrix{I_{2} & \cr
                              & R_{s}}.
\ee
\noindent
Bars denote either an  
$SO(10)$ or an $SO(5,5)$ representation acting on 
the multiplet bases of Ref.\cite{seva} depending on the 
metric involved.  
 Then we see that the equation  
\be
(\bar{R}_{s}^{32})(\bar{\Gamma}^{s}_{c})^{m}(\bar{R}_{s}^{32})^{-1}
=(\bar{\Omega}^{-1})^{m}_{n}(\bar{\Gamma}^{s}_{c})^{n}
\ee
\noindent
is satisfied consistently
when we use eqns. (5) and (8) and 
 $\bar{R}_{s}^{32} = I_{2}\,
\otimes{\bar{R}_{s}(\Omega)}$. The relation (10) is a generalisation
of our familiar relation $S^{-1}\, \gamma^{\mu}\,S = g^{\mu \nu}\,
\gamma_{\nu}$ in 4-d quantum mechanics\cite{quantum}. 
As an aside we note that the relation (10) gives back,
\be
R_{c}(\Omega^{0}) = \pmatrix{ & &I_{2} &0  \cr
                              & & 0 &-\sigma_{3} \cr      
                              I_{2}&0 & &    \cr      
                              0 &-\sigma_{3} & &  },\quad
R_{c}(\Omega^{\star}) = \pmatrix{ & & I_{2}&0 \cr
                                  & & 0 &-I_{2} \cr
                                 I_{2} &0 & & \cr
                                     0 &-I_{2}& & }
\ee
\noindent
In the relation (10) 
\be
(\bar{\Gamma}^{s}_{c})^{m} = \bar{\eta}^{m}_{n}\,(\Gamma^{s}_{c})^{n},\nonumber
\quad\quad(\Gamma^{s}_{c})^{m} = (I_{2} \otimes{U^{s}_{16}})\Gamma^{m}_{W}
(I_{2} \otimes{U^{s}_{16}})^{T},\\
\ee
with
\bea
\bar{\eta} = (1/\sqrt{2})\,\pmatrix{1 & -1 & \cr
                                         1 &  1 & \cr
                                           &    & I_{8}}.\nonumber
\eea
\noindent
They satisfy Clifford algebra:
\bea
\{(\bar{\Gamma}^{s}_{c})^{m},(\bar{\Gamma}^{s}_{c})^{n}\} = 2\bar{L}_{c}^{mn},
\nonumber\quad
\{(\Gamma^{s}_{c})^{m},(\Gamma^{s}_{c})^{n}\} = 2I_{10}^{mn},\nonumber\quad
\{\Gamma^{m}_{W},\Gamma^{n}_{W}\} = 2I_{10}^{mn},\nonumber\\
\eea
with
$\bar{L}_{c} = \pmatrix{\sigma_{1} &  \cr
                              & I_{8}}$.
$U^{s}_{16}$ in eq.(12) is given by 
\be
U^{s}_{16} = \pmatrix{A & -A.\epsilon\,\otimes{\epsilon}\,\otimes{\epsilon}\cr
                   A.\sigma\otimes{I_{2}}\otimes{\epsilon} & 
                   A.\sigma_{3}\otimes{\epsilon}\,\otimes{I_{2}}},
A = \pmatrix{1&0&0&-1&0 &-1 &-1 & 0\cr             
             0&1&1&0 &1&0&0&-1\cr
             1&0&0&1&0 &1 &-1 & 0\cr
             0&-1&1&0 &1&0&0&1\cr
             1&0&0&-1&0 &1 &1 & 0\cr
             0&1&1&0 &-1&0&0&1\cr                    
             1&0&0&1&0 &-1 &1 & 0\cr             
             0&1&-1&0 &1&0&0&1},
\ee    
$\Gamma_{W}$'s in eq.(12) are gamma 
matrices in a Weyl basis and are given explicitly as,
\bea 
(\Gamma_{W})^{0}=\sigma_{2}\otimes{I_{16}},& &
(\gamma_{8})^{1}=\epsilon\otimes\epsilon\otimes\epsilon,\nonumber\\
(\Gamma_{W})^{1-5}=\sigma_{1}\otimes\epsilon
\otimes(\gamma_{8})^{1-5},& & 
(\gamma_{8})^{2}=1\otimes\sigma_{1}\otimes\epsilon,\nonumber\\  
(\Gamma_{W})^{6-7}=\sigma_{1}\otimes\epsilon
\otimes(\gamma_{8})^{6-7},& & 
(\gamma_{8})^{3}=1\otimes\sigma_{3}\otimes\epsilon,\nonumber\\
(\Gamma_{W})^{8}=-\sigma_{1}\otimes\sigma_{1}\otimes{I_{8}},& & 
(\gamma_{8})^{4}=\sigma_{1}\otimes\epsilon\otimes{1},\nonumber\\
(\Gamma_{W})^{9}=\sigma_{1}\otimes\sigma_{3}\otimes{I_{8}},& & 
(\gamma_{8})^{5}=\sigma_{3}\otimes\epsilon\otimes{1},\nonumber\\
(\Gamma_{W})^{11}=-\sigma_{3}\otimes{I_{16}},& & 
(\gamma_{8})^{6}=\epsilon\otimes{1}\otimes\sigma_{1},\nonumber\\  
                & & (\gamma_{8})^{7}=\epsilon\otimes{1}\otimes\sigma_{3}
\nonumber
\eea
with $\epsilon=i\sigma_{2}$.
\noindent
A set of $SO(1,9)$ gamma matrix 
representations in Dirac basis could be found in Ref.\cite{corn,GSW}.

We note that once we replace 
$( \bar{\Gamma}^{s}_{c}, \Gamma^{s}_{c},\bar{L}_{c} )$ 
by the corresponding noncompact ones
 $( \bar{\Gamma}^{s}, \Gamma^{s},\bar{L} )$ 
in eqns. ( 10, 12, 13 ), 
eq.(10) holds true, if we substitute $\bar{\Omega}$ 
and $\bar{R}^{s}(\bar{\Omega})$ of eq. (9) by any
 $T$-duality element embedded in $SO(5,5)$ or the $U$-duality
element $\bar{\Omega}_{0}$, of Ref.\cite{seva}.
 $\Gamma^{s}$'s are related to $\Gamma^{s}_{c}$'s as:\quad
$(\Gamma^{s})^{j} = i (\Gamma^{s}_{c})^{j}$ for j=1,...,5;
otherwise,they are same. Here, $\bar{L} = \pmatrix{\sigma_{1}&  \cr
                                                       &\bar{L}_{8}}$,
\quad$\bar{L}_{8}=\pmatrix{-I_{4} & \cr
                                  & I_{4}}$.

\par
Now let us come back to eqns. (5) and (8).
We find that,
\noindent
\bea
\Omega(Z_{2}^{0})\,\tilde{L}\,
\Omega(Z_{2}^{0})^{T}=-\tilde{L},\nonumber\quad
det\,\Omega(Z_{2}^{0})=-1\nonumber\\
\noindent
\Omega(Z_{2}^{\star})\,\tilde{L}\,\Omega(Z_{2}^{\star})^{T}
=-\tilde{L},\quad
det\,\Omega(Z_{2}^{\star})=-1.
\eea
Eq.(15) implies that
 $\Omega(Z_{2}^{0})$ and $\Omega(Z_{2}^{\star})$
do not correspond to $SO(5,5)$ transformations.\\
To investigate the relation between the two symmetries, we
note that
$Z_{2}^{0}$ acting on $(Q ,P ,\tilde{M})$ 
and $Z_{2}^{\star}$ acting on $(Q^{'},P^{'},\tilde{M}^{'})$
separately are symmetries of action where Q and $\rm{Q}^{'}$ etc.
act as dummy symbols for multiplets of field variables
appearing in the action.
Now to establish their relationship let us assume that 
the action of $Z_{2}^{0}$ on 
$(Q,P,\tilde{M})$ and that of $Z_{2}^{\star}$ on 
$(Q^{'},P^{'},\tilde{M}^{'})$ are related
by a symmetry element belonging to the equation of motion 
and see that this assumption is consistent or, in other words,
 we do not face
any contradiction. Now we see that this assumption enables
us to write:
\noindent
\bea
Q^{'}=\tilde{\Omega}Q,\nonumber \quad
P^{'}=\tilde{R_{s}}\,P,\nonumber\quad
\tilde{M}^{'}=\tilde{\Omega}\,\tilde{M}\,\tilde{\Omega}^{T},\nonumber\\
\noindent
\Omega(Z_{2}^{\star})
=\tilde{\Omega}\,\Omega(Z_{2}^{0})\,\tilde{\Omega}^{-1},\nonumber\quad
\tilde{R_{s}}(Z_{2}^{\star})
=\tilde{R_{s}}\tilde{R_{s}}(Z_{2}^{0})\tilde{R_{s}}^{-1},\nonumber\\
\eea
where, $\tilde{\Omega}$ and $\tilde{R_{s}}$ are relating matrix 
transformations of $Z_{2}^{0}$ and $Z_{2}^{\star}$ on vector
 and spinor multiplets respectively.
We find the explicit form of $\tilde{\Omega}$, $\tilde{R_{s}}$  as:
\noindent
\bea
\tilde{\Omega}=\pmatrix{ \sigma_{1}& & & & \cr
                                            &0&0&I_{3}&0 \cr
                                            &0&1&0 &0 \cr
                                            &I_{3}&0&0&0 \cr
                                            &0&0&0&1 },
\eea
\noindent
\bea
\tilde{R_{s}}= - \pmatrix{0&I_{8}\cr
                  I_{8}&0},
\eea
\noindent       
where $\tilde{\Omega}$ satisfies
 $\tilde{\Omega}\tilde{L}\tilde{\Omega}^{T}
=\tilde{L}$
\,and\, $det\tilde{\Omega} = 1$. This implies that
 $\tilde{\Omega}$ in eq.(17) is 
an $SO(5,5)$ vector transformation. It further
implies that $\it{Q, \tilde{M}}$ and 
$\it{Q^{'}, \tilde{M}^{'}}$ satisfy the same 
equation of motion. So here, we do not face 
any contradiction. Then our task is to 
show that $\tilde{R}_{s}$ also belongs to $U$-duality
symmetry $SO(5,5)$ and it corresponds to the 
spinor representation of the
element whose vector representation is $\tilde{\Omega}$.\\
 
Now let us start by making some observation about the
spinor multiplets $P_{\bar\mu}$ and
$P_{\bar{\mu}}^{'}$. 
They are in 16-dimensional forms
making it essential for us to consider 
our $SO(5,5)$ spinor representation 
in a Weyl basis where all generators 
in the 32-dimensional 
 spinor representation 
  reduce \cite{corn,wett}
to two 16-dimensional blocks 
along the diagonal. The entries of the two blocks are either
 same or differ by an overall minus sign. 
Now we can find any Weyl basis for gamma
matrix representation which keeps the form of
$\it{P_{\bar{\mu}}}$ unchanged and satisfy
the relation 
\be
(\tilde{R_{s}}^{32})
\tilde{\Gamma}^{m}(\tilde{R_{s}}^{32})^{-1}
=(\tilde{\Omega}^{-1})^{m}_{n}\tilde{\Gamma}^{n}.
\ee
\noindent
where in (19),
 $\tilde{R_{s}}^{32}\,=\,I_{2}\,\otimes\,\tilde{R_{s}}$\quad
i.e. $\tilde{R}_{s}^{32}=\tilde{R_{s}}(\tilde{\Omega})$,
thereby proving that $\tilde{R}_{s}$ in eq.(18)
corresponds to the
spinor representation of the element whose 
vector representation is $\tilde{\Omega}$.
\par 
Now instead of choosing an arbitrary
gamma matrix representation in this way
let us go about in a bit roundabout path to 
determine one representation which will be useful 
for us when we will be considering simultaneous
modding out by `Orientifold' pair of discrete
symmetries and $T$-duality symmetries of Ref.\cite{seva}.
Let us start by noting that
 $\tilde{\Omega}=\tilde{\Omega}_{0}\,\tilde{L}^{10}$,
(throughout the paper tilde refers to objects taken
in metric $\tilde{L}$ of Ref.\cite{seva})
 with
\be
\tilde{\Omega}_{0}=\pmatrix{\sigma_{1} & & & & \cr
                                   &I_{3}&0&0&0 \cr
                                   &0&0&0&1 \cr
                                   &0&0&I_{3}&0 \cr
                                   &0&1&0&0 },\nonumber\,
\tilde{L}^{10}=\pmatrix{ I_{2}&  \cr
                       &\tilde{L}_{8} }.\nonumber
\ee
In eqn.(20), $\tilde{\Omega}_{0}$ is obtained from  
 $\bar{\Omega}_{0}$ in ref.\cite{seva} as:
\noindent
\bea
\tilde{\Omega}_{0}=\eta^{10}\,
\bar{\Omega}_{0}\,(\eta^{10})^{T},\nonumber\quad
\eta^{10}=
\pmatrix{I_{2}&  \cr
              &\eta^{'}},\nonumber\quad
\eta^{'}=1/\sqrt{2}\pmatrix{I_{4}&I_{4}\cr
                      -I_{4}&I_{4}},\nonumber\\
\eea
where  $\eta^{10}$ takes into account the metric change
from ref.\cite{seva} to ref.\cite{Kumar}.
Here $\tilde{L}^{10}$ is a $T$-duality element.
It is obtained from $\bar{L}^{10}$ in the same
way as $\tilde{\Omega}_{0}$ from $\bar{\Omega}_0$.
Now we obtain $\tilde{R}_{s}$ in eq.(18) from 
$\bar{R}_{s}(\bar{\Omega}_{0}\,\bar{L}^{10})$ 
as:
\noindent
\be
\tilde{R_{s}} = (V_{16}\,\eta^{16})\,
\bar{R}_{s}(\bar{\Omega}_{0}\,\bar{L}^{10})\,
(V_{16}\,\eta^{16})^{T},
\ee
\noindent
where,
\bea
\bar{R}_{s}(\Omega_{0}) = \pmatrix{0 & I_{8}\cr
                                   -I_{8}&0},\nonumber\quad
\bar{R}_{s}(\bar{L}^{10}) = \pmatrix{\bar{L}_{8} & \cr
                                                 &-\bar{L}_{8}}.
\eea
\noindent
So the spinor multiplet in Ref.\cite{Kumar}
is related to that in Ref.\cite{seva} as:
$P_{\bar{\mu}} = (V_{16}\,\eta^{16})\,\bar{P}_{\bar{\mu}}$. 
As a result $\tilde{R}_{s}$ relates two spinor multiplets 
which satisfy the same equation of motion as
$\bar{R}_{s}(\bar{\Omega}_{0})$ and $\bar{R}_{s}(\bar{L}^{10})$  
do. Infact, we verify that
$\tilde{R}_{s}$ and $\tilde{\Omega}$ 
refer to 
same $SO(5,5)$ element 
as they 
satisfy
\noindent
\be
(\tilde{R}_{s}^{32})
\tilde{\Gamma}^{m}(\tilde{R}_{s}^{32})^{-1}
=(\tilde{\Omega}^{-1})^{m}_{n}\tilde{\Gamma}^{n}.
\ee
\noindent
This is consistent with our initial assumption. In eq.(23),
 $\tilde{R}_{s}^{32}\,=\,I_{2}\,\otimes\,\tilde{R}_{s}$
and
\noindent
\bea
\tilde{\Gamma}^{m} = \tilde{\eta}^{m}_{n}\Gamma^{n},\nonumber\quad
\Gamma^{m} = (V\eta^{32})(\Gamma^{s})^{m}(V\eta^{32})^{T},\nonumber\quad
\eta^{32}=I_{2}\otimes{\eta^{16}},\nonumber\quad
 V=I_{2}\otimes{V_{16}},\nonumber\\
\tilde{\eta}=(1/\sqrt{2})\pmatrix{
1&-1&0&0\cr
1&1&0&0\cr
0&0&I_{4}&I_{4}\cr
0&0&-I_{4}&I_{4}},\nonumber\quad
\eta^{16} =\pmatrix{\eta^{'} & \cr
                           & \eta^{'}},\nonumber\quad
V_{16} = \pmatrix{I_{4}& & & \cr
                     &I_{4}& & \cr
                     &     &0 &I_{4}\cr
                     &     &I_{4}&0 }.\nonumber\\
\eea
\noindent
$\Gamma$'s satisfy the Clifford algebra:
\be
\{\tilde{\Gamma}^{m},\tilde{\Gamma}^{n}\}
=2\tilde{L}^{mn},\quad
\{\Gamma_{W}^{m},\Gamma_{W}^{n}\}=2l^{mn},
\ee
 where, $l^{mn}=diag(1,-I_{5},I_{4})$.  
It is interesting to note here that
we do not face any contradiction to the assumption that
the two discrete symmetries
are related and we find that they are 
related by an $SO(5,5)$ element as well as their explicit
form.
Hence we have therefore found an explicit relationship between 
the symmetries  $Z_2^0$ and $Z_2^*$. 

\par
Now we make connection between 
some of the results of
Ref.\cite{seva} and Ref.\cite{Kumar}.
For that let us work in metric $\tilde{L}$ of Ref.\cite{Kumar} 
 and spinor multiplet basis, $V_{16}^{-1}\,P_{\bar{\mu}}$. 
So we will operate on 
objects associated
with spinor representations of 
Ref.\cite{Kumar} like 
 $Z_{2}^{0}$  and  $Z_{2}^{\star}$ etc. by $V_{16}^{-1}$;
and use superscript `s' to denote new ones.
We will need apply $\eta^{10}$ and $\eta^{16}$ matrices
on objects associated with
 vector and spinor representations respectively
of Ref.\cite{seva}.
Now we show that for specific $T$-duality elements,
$\tilde{\Omega}$ relating orientifold
pairs of Ref.\cite{Kumar} 
and $\tilde{\Omega}_{0}$ relating
$T$-dual pairs of Ref.\cite{seva} are equivalent.
For that let us consider a  
$T$-duality element embedded in $SO(5,5)$ of Ref.\cite{seva}
in the metric $\tilde{L}$
 and spinor multiplet basis, $V_{16}^{-1}\,P_{\bar{\mu}}$. 
Let us denote its vector and spinor representations as
 $\tilde{\Omega}^{t}$,\quad
 $\tilde{R_{s}}^{t}$. We 
use superscript `t' 
to refer $T$-duality.
We see that this element 
is given in terms of 
$T$-duality
element of Ref.\cite{seva}
in the metric $\bar{L}$  as:
\be
\tilde{\Omega}^{t}=\pmatrix{I_{2}& 0 \cr
                             0&\eta^{'}R_{s}(\Omega)(\eta^{'})^{T}},
\tilde{R_{s}}^{t}=\pmatrix{\eta^{'}R_{c}(\Omega)(\eta^{'})^{T}& \cr
                                    &\eta^{'}\Omega(\eta^{'})^{T}}.
\ee
\noindent
Then applying $\tilde{\Omega}$ 
and $\tilde{R_{s}}^{s}$  on $\tilde{\Omega}^{t}$\,$\rm{and}$\,
$\tilde{R_{s}}^{t}$ respectively we get the
duals (denoted as primed):
\noindent 
\bea 
(\tilde{\Omega}^{t})^{'}&=&\tilde{\Omega}\tilde{\Omega}^{t}
\tilde{\Omega}^{-1},\nonumber\\
&=&\eta^{10}\bar{\Omega}_{0}
\pmatrix{I_{2}& \cr
              &(R_{s}(\Omega)^{T})^{-1}}
\bar{\Omega}_{0}
(\eta^{10})^{T},\\
\noindent
(\tilde{R_{s}}^{t})^{'}&=&\tilde{R_{s}}^{s}\tilde{R_{s}}^{t}
(\tilde{R_{s}}^{s})^{-1},\nonumber\\
&=&\eta^{16}
R_{s}(\Omega_{0})
\pmatrix{(R_{c}(\Omega)^{T})^{-1}& \cr
                                 &(\Omega^{T})^{-1}}
R_{s}(\Omega_{0})^{-1}
(\eta^{16})^{T}.
\eea
\noindent
In eqns. (27) and (28) we have used eqns. (2.4), (2.6), (2.15) and (2.30)
of Ref.\cite{seva}.
Now for all the examples of Ref.\cite{seva}, the matrices $
\Omega, R_{c}(\Omega), R_{s}(\Omega))$ are 
orthogonal matrices.  
Hence for them action of
$(\tilde{\Omega}_{0},\tilde{R_{s}}(\Omega_{0}))\quad
  $\rm{and}$
\quad (\tilde{\Omega},\tilde{R_{s}}(\Omega))$ are same,
or 
in other words $\tilde{\Omega}$ and $\tilde{\Omega}_{0}$
are equivalent.\\

Now let us discuss the constructions of models by
simultaneous projections by the $T$-duality elements of
\cite{seva} together with $Z_2^0$ or $Z_2^*$.
Here we note in the context of eq.(26) that the matrices
$\Omega,R_{c}(\Omega)$ and $R_{s}(\Omega)$ satisfy
$\eta^{'}\,(R_{s}(\Omega),R_{c}(\Omega),\Omega)\,(\eta^{'})^{T}
=(R_{s}(\Omega),R_{c}(\Omega),\Omega)$,
when  $\Omega,R_{c}(\Omega),R_{s}(\Omega)$ are 
of the form of 8-dimensional matrices
$\pmatrix{A& \cr
          &A}$, where $A$ denotes a 4-dimensional diagonal block. 
 Then $\tilde{\Omega}^{t}$ and $\bar{\Omega}^{t}$
are identical and so are $\tilde{R}_{s}^{t}$ and $\bar{R}_{s}^{t}$.
For the examples of Ref.\cite{seva} the
same thing happens with the dual elements also. 
One such $T$-duality element in 
Ref.\cite{seva} is $\Omega = (\pi,0,\pi,0)$.
These pairs of
 $T$-duality elements of Ref.\cite{seva}\quad are all
 diagonal. Again $V^{-1}\,\Omega(Z_{2}^{0})\,V$ and
$V^{-1}\,\Omega(Z_{2}^{\star})\,V$ are also diagonal.
Hence we get unambiguously fields from 
unprimed multiplets 
which are invariant under both
 $T$-duality element, discussed above 
 and $V^{-1}\,\Omega(Z_{2}^{0})\,V$.
In the same way we get from
primed multiplets simultaneous
invariants under 
the dual $T$-duality element and 
$V^{-1}\,\Omega(Z_{2}^{\star})\,V$ on the other side 
giving us
dual pair of models. 
 By this process of intersection we get  
 models in four, three or two dimensions respectively. One such model
in four dimension has been given in \cite{Kumar}.\\

To conclude,
we have verified that for construction of typeIIA dual pairs through
orientifolding, an explicit $U$-duality relationship for fields
in the vector as well as spinor representation can be constructed.
Unlike the pairing of type IIA on K3,with heterotic string on $T_{4}$,
\cite{gao} this relationship has been proven through the construction of
appropriate gamma matrices.
We can now follow Ref.\cite{sen2} to test duality of such pair of
models. It will be interesting to see how such models fit 
in the framework of $M$ and $F$ theories 
\cite{mth2,vafa,Morrison,sen}.

I would like to thank Ashoke Sen for several crucial
 suggestions and discussions. I am greatly indebted to
Alok Kumar for guidence 
and encouragement throughout the course of this work. 
I would also like to thank Dileep Jatkar,
 Rajesh Parwani for discussions, Harvendra Singh,
 Koushik Roy for critical reading of the manuscript
 and Mehata Research Institute,
Allahabad,for hospitality where this work was started. 


\end{document}